\documentclass[floatfix,twocolumn,showkeys,showpacs,prb,amsmath,amssymb]{revtex4}

\usepackage{epsfig}
\usepackage{dcolumn}
\usepackage{bm}
\usepackage{color}
\usepackage{moreverb}  

\begin{document}


\title{Enhancing structure relaxations for first-principles codes:\\an approximate Hessian approach}

\author{James M Rondinelli}%
	\altaffiliation[Present address: ]{Materials Department, University of California, Santa Barbara, California, 93106-5050, USA}
	\email{rondo@mrl.ucsb.edu}
\author{Bin Deng}%
	\email{b-deng@northwestern.edu}
\author{Laurence D Marks}
 \email[Corresponding author: ]{L-marks@northwestern.edu}
\affiliation{%
Department of Materials Science and Engineering, Northwestern University, 2220 N Campus Drive, Evanston, Illinois, 60208-3108, USA}%

\date[Submitted: ]{Aug.\ 15, 2006; Revised: \today}
\preprint{COMMAT-D-06-00288}

\begin{abstract}
We present a method for improving the speed of geometry relaxation by using a harmonic approximation for the interaction potential between nearest neighbor atoms to construct an initial Hessian estimate. The model is quite robust, and yields approximately a 30\% or better reduction in the number of calculations compared to an optimized diagonal initialization. Convergence with this initializer approaches the speed of a converged BFGS Hessian, therefore it is close to the best that can be achieved. Hessian preconditioning is discussed, and it is found that a compromise between an average condition number and a narrow distribution in eigenvalues produces the best optimization.
\end{abstract}

\maketitle

\section{\label{sec:introduction}Introduction}

In many cases the slowest step in a density functional calculation (DFT) or other \emph{ab initio} calculations is finding the optimal atomic positions which minimize the total energy. With older minimization approaches, such as the conjugate gradient method, the number of evaluations scales proportionally with the systeFcommm size. More powerful are quasi-Newton methods, in particular the Broyden-Fletcher-Goldfarb-Shanno (BFGS) method, which can show quadratic convergence provided that breakdowns of the curvature condition (discussed later) are protected against. Essential to the quasi-Newton methods are estimates for the gradient and curvature of the potential energy surface; the latter being stored in a matrix commonly referred to as the Hessian, which contains all second derivatives (or atomic force constants). The classic BFGS method uses a simple diagonal matrix as the initial Hessian estimate, perhaps with the initial diagonal term using the Shanno-Phua scaling \cite{Shanno-Phua}; see also the discussion by Nocedal and Wright \cite{nocedal:wright:1999}. In principle one could achieve far better convergence by some appropriate choice of the initial Hessian estimate, as suggested by some recent analysis \cite{PhysRevB.64.161102,fernandez-serra:100101,mostofi:8842,nemeth:2877}.

In this paper, we detail an approach for improving on the estimate of the starting Hessian, using a harmonic potential describing the interactions between nearest neighbor atoms. We find that it is important to combine this with a diagonal component plus an appropriate scaling term. Slightly unexpectedly, what turns out to be important is a balance between making the initial Hessian estimate replicate that of the true problem and keeping the condition number of the estimate small. 

The structure of this note is as follows. First, we briefly review conventional optimization methods (Section~\ref{sec:optmethods}), with some comments about how they might be improved for density functional theory (DFT) calculations. Second, we outline the algorithm for generating the Hessian estimate and implementing it into the all-electron (linearized) augmented-plane wave + local orbitals (L/APW+lo) package WIEN2k \cite{REF:Bla2001} (Section~\ref{sec:implementation}).  The robustness of the program is tested by performing geometry relaxations for various classes of materials (Section~\ref{sec:results}). Finally, we conclude with a discussion on the importance of Hessian preconditioning, and we propose a general scheme for resolving these problems.

\section{\label{sec:optmethods}Optimization Methods}
At the heart of quasi-Newton methods is an expansion of the energy in the form
\begin{equation} \label{eqt:energy}
E^\dag   = E + {\mathbf{g}}^T {\mathbf{s}} + \frac{1}{2}{\mathbf{s}}^T {\mathbf{Hs}}
\end{equation}
where $E^\dagger$ is the predicted energy, $E$  and $\mathbf{g}$  are the energy and gradient for a step $\mathbf{s}$  from the current state, and $\mathbf{H}$ is the Hessian matrix.  The optimum step can be obtained directly in principle as
\begin{equation}
{\mathbf{s}} =  - {\mathbf{H}}^{ - 1} {\mathbf{g}}
\end{equation}
assuming that the Hessian is known. The concept of a quasi-Newton method is to calculate an approximation to the Hessian (or its inverse depending upon the exact method used) from previous gradient information. The most successful approaches use what are called secant methods \cite{schlegel:1995}, in particular the Broyden-Flecher-Goldfarb-Shanno (BFGS) method \cite{C.G.BROYDEN09011970,R.Fletcher03011970,Goldfarb:1970:FVM,Shanno:1970}. The most important contribution from these minimization algorithms is the use of Hessian updating techniques, which allow for the collection of more information about the energy surface. In general, after each cycle the Hessian is updated during the minimum search until the convergence criterion is satisfied.  It is important to recognize that convergence can be achieved without ever reaching the true Hessian, which suggests that the efficiency of the structure relaxation depends  on both the starting geometry and the initial conditioning of the Hessian estimate (discussed later). In fact, the true Hessian of the problem is not always the optimal one, and a compromise between conditioning and accuracy is much more desirable for optimization problems; as Baali has shown much of the success of quasi-Newton methods relies on self-scaling algorithms \cite{baali1,baali2}. The first estimate for the Hessian is usually a unitary matrix, although this is not required if physical knowledge of the system is available. For instance, in an earlier version of the WIEN2k code \cite{REF:Bla2001} an estimate of the bonding force constants and atom multiplicities was used for the initial diagonal elements---this worked much better than a simple constant. As we will see, one can do better than this.

The mathematics behind the secant method is that a typical iteration for the minimization of $f(x)$ is given by the form
\begin{equation}
x_{k + 1}  = x_k  + \alpha _k d_k 
\end{equation}
where $d_k  =  - {\mathbf{B}}^{-1}_{k} \nabla f\left( {x_k } \right)$ and ${\mathbf{B}}_k $
 is the approximation for the true Hessian that is updated and the step size $\alpha _k $ is chosen by a line search or a trust-region method (as here) \cite{CONN:2000,Flet80,Kelley:1999}.

For any two consecutive iterations, $x_k$ and $x_{k + 1}$, with their gradients, $\nabla f\left( {x_k } \right)$  and $\nabla f\left( {x_{k + 1} } \right)$ information about the curvature of the surface (the Hessian) is known since
\begin{equation}
\left[ {\nabla f\left( {x_{k + 1} } \right) - \nabla f\left( {x_k } \right)} \right] \approx {\mathbf{B}}_{k + 1} \left[ {x_{k + 1}  - x_k } \right]
\end{equation}
writing ${\mathbf{s}}_k  = x_{k + 1}  - x_k $ and ${\mathbf{q}}_k  = \nabla f\left( {x_{k + 1} } \right) - \nabla f\left( {x_k } \right)$, this can be rewritten as
\begin{equation} \label{eqn:secant}
{\mathbf{q}}_k  = {\mathbf{B}}_{k + 1} {\mathbf{s}}_k .
\end{equation}
The expression given in Eq.~\ref{eqn:secant}  is known as the secant equation. An important constraint is that ${\mathbf{B}}_{k + 1}$ needs to be positive definite for the step to be downhill. Multiplying Eq.~\ref{eqn:secant} on the left by ${\mathbf{s}}_k$ yields what is called the curvature condition ${\mathbf{s}}_k  \cdot {\mathbf{q}}_k  > 0$. This is equivalent to the geometric interpretation that over the step length the object function has positive curvature (i.e.,~the step is taken in a lower energy direction). When this condition is satisfied, Eq.~\ref{eqn:secant} will always have a solution and the BFGS update
\begin{equation} \label{eqn:BFGS}
{\mathbf{B}}_{k + 1} = {\mathbf{B}}_k  + \Delta {\mathbf{B}}_k,\quad \Delta {\mathbf{B}}_k = \frac{{{\mathbf{q}}_k {\mathbf{q}}_k^T }}
{{{\mathbf{q}}_k^T {\mathbf{s}}_k }} - \frac{{{\mathbf{B}}_k {\mathbf{s}}_k {\mathbf{s}}_k^T {\mathbf{B}}_k }}
{{{\mathbf{s}}_k^T {\mathbf{B}}_k {\mathbf{s}}_k }}
\end{equation}
will maintain a positive definite approximation to the Hessian. 

It is worth mentioning that the curvature condition does not always hold, so it must be explicitly enforced otherwise the BFGS method can fail completely; this is one of the weaknesses of these updating methods. This often occurs when the character of the Hessian changes substantially during the course of the minimization, which is more likely to occur if one starts far from the minimum. Fortunately, the BFGS update is rather well behaved, in that the Hessian estimate will tend to correct itself in a few steps, as compared to other approaches \cite{nocedal:wright:1999}. Three conventional techniques exist for handling the case when the curvature condition fails:
\begin{enumerate}
	\item The calculations are restarted from the current position with a diagonal initial estimate.
	\item	A skipping strategy is employed on the BFGS update $(  {\mathbf{B}}_{k + 1}  = {\mathbf{B}}_k)$.
	\item The use of a revised (damped) BFGS update \cite{nocedal:wright:1999} which modifies the definition of ${\mathbf{q}}_k $.
\end{enumerate}
For the first case, any important curvature information is lost and previous steps are essentially wasted. The second technique allows one to incorporate the curvature information at previous iterations. However, it requires careful control, and too many updates may be skipped resulting in further loss of curvature information. (The limited memory method \cite{LBFGS, nocedal:wright:1999} can do this better because it skips steps far from the current location.) The particular code we employed used the third method where the scalar ${\mathbf{t}}_k$ is defined by ${\mathbf{t}}_k  = \left( {0.2} \right){\mathbf{s}}_k^T {\mathbf{B}}_k {\mathbf{s}}_k  > 0$ and ${\mathbf{u}}_k  = \theta _k {\mathbf{q}}_k  + \left( {1 - \theta _k } \right){\mathbf{B}}_k {\mathbf{s}}_k $ for
\[
\theta _k  =
\begin{cases}
   1, & {\text{ if }} \quad{\mathbf{s}}_k^T {\mathbf{q}}_k  \geqslant {\mathbf{t}}_k \\
   0.8\frac{{{\mathbf{s}}_k^T {\mathbf{B}}_k {\mathbf{s}}_k }} 
{{{\mathbf{s}}_k^T {\mathbf{B}}_k {\mathbf{s}}_k  - {\mathbf{s}}_k^T {\mathbf{q}}_k }}, & {\text{ if }} \quad {\mathbf{s}}_k^T {\mathbf{q}}_k  < {\mathbf{t}}_k 
\end{cases}.
\]
The BFGS update is then given just as in Eq.~\ref{eqn:BFGS}, with ${\mathbf{q}}_k $ replaced by ${\mathbf{u}}_k $. This formulation enforces the curvature condition, and allows for an interpolation between the unmodified Hessian update $({\theta _k  = 1})$ and the Hessian at the current iterate. As a consequence, every step contributes to defining the curvature, and no steps are wasted

Much of the previous discussion has been concerned with the BFGS update, but selection of the step size $\alpha _k $ (and direction) merits attention. In the code we have (\textsc{dmng.f} from NETLIB with some minor changes) the entire BFGS update is wrapped within a trust-region algorithm which is used to calculate the best step to take based on a quadratic model of the objection function (the PES). While line search methods can be used, for a DFT problem where the gradient comes essentially for free, the most efficient approach is a trust-region algorithm \cite{355966}. In this method information about the object function $f$ is collected to construct a quadratic model function $\mathcal{L}$ that is said to adequately sample $f$ in the neighborhood of the current iterate. The model function 
\[
\mathcal{L}(\mathbf{s}_{k+1})=f({s}_{k})+\mathbf{g}^T_k\mathbf{s}_{k+1}+\frac{1}{2}\mathbf{s}_{k+1}^T\mathbf{B}_k\mathbf{s}_{k+1}
\]
uses the current estimate of the Hessian and imposes an additional constraint on the step length, $\left\| {{\mathbf{s}}_{k + 1} } \right\| \leqslant R$ where $R$ is the trust region radius. The step size which minimizes $f$ is then chosen such that it sufficiently minimizes $\mathcal{L}$ over the trust region; the radius $R$ is then adjusted iteration to iteration according to how well the step reduced the function with respect to the predicted reduction value determined from the step size (a so called effectiveness measure) \cite{nocedal:wright:1999}. Therefore, if a poor step is taken, the radius is decreased, until the current Hessian approximation is good enough, and then it is subsequently expanded. Compared to line search methods, this approach may not give the best improvement per direction, but often will be faster in terms of the nett improvement per function evaluation.

\begin{figure}
	\centering
		\includegraphics[width=0.40\textwidth]{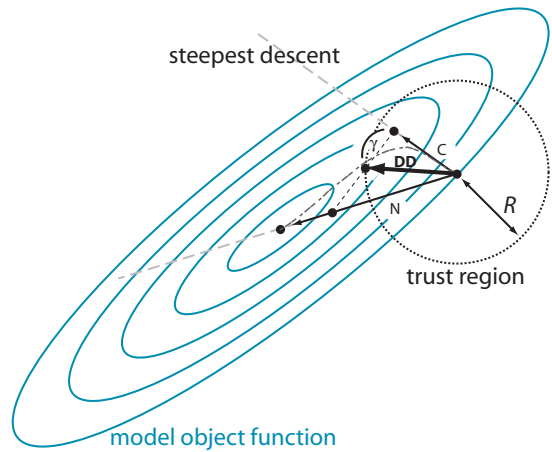}
		\caption{\label{fig:steps}(color online)~Schematic illustration of the principle steps used in the BFGS algorithm. The curved trajectory (dashed-dotted line) corresponds to the optimal path to the minimizer of the object function. The step is selected based on the trust region (dotted circle) of radius $R$ and how well it minimizes the objection function (elliptical contours). The Cauchy step (C) is perpendicular to the contour lines of the object function at the current position, and the full Newton step (N) is shown along the gradient of the object function. The double dogleg step (DD) is shown to be biased towards the Newton direction according to the implementation of the Dennis and Mei algorithm (cf.\ Ref.\ \cite{Dennis:1979}).}
\end{figure}

The routine in our minimization has the added feature of using an adaptive trust region method, in that it switches between different models in order to determine the optimal step size \cite{Dennis:1979,Dennis:1981:ANL}. The algorithm first calculates what it believes is an appropriate step size (such that the length of the step is less than the radius of the trust region). It is unusual to calculate the exact trust region step, so approximate trial steps are found which approximate the solution in this region. For each iteration the algorithm computes the step size as the linear combination of the steepest descent direction and a quasi-Newton search direction. Different step types are then chosen (Fig.\ \ref{fig:steps}), the main ones being:
\begin{enumerate}
	\item	A restricted Cauchy step ($\mathbf{s}_{RC}=-R\frac{\mathbf{g}_{k}}{\left\|\mathbf{g}_{k}\right\|}$) if the trust radius $R$ is smaller than or equal to the Cauchy step ($\mathbf{s}_{C}=-\frac{\mathbf{g}_{k}^T\mathbf{g}_{k}}{\mathbf{g}_{k}^T\mathbf{B}_{k}\mathbf{g}_{k}}\mathbf{g}_{k}$). The Cauchy step is taken along the gradient of the quadratic model of the object function and is of a length defined by the minimizer of that function. 
	\item	A full Newton step ($\mathbf{s}_{N}=-\mathbf{B}_{k}^{-1}\mathbf{g}_{k}$) is taken in length and direction if the trust region allows it (large enough), otherwise a step in the Newton direction of a limited length is taken, such that it satisfies the constraint of $R$.
  \item A double dogleg step ($\mathbf{s}_{DD}=\mathbf{s}_{C}+\gamma_k(\mathbf{s}_{N}-\mathbf{s}_{C})$) if the trust radius is between the Cauchy and Newton steps. The $\gamma_k$ parameter (see Fig.\ \ref{fig:steps}) describes the length and direction between the Cauchy point and the intersection of the trust region. The double dogleg may therefore be seen as a compromise between the Cauchy step and Newton step, whereby the direction and length is given by the line connecting these steps through the intersection of $R$.
\end{enumerate}

This approach is used since it provides a stronger bias toward the Newton step direction, as an attempt to accelerate the optimization. The purpose of the Cauchy step is just to minimize the local model over a space that is known to be well defined in the steepest descent direction. Since steepest descent directions do not always provide the best minimization, alternative candidate steps (and directions) are evaluated.
Conventional BFGS methods as outlined above work very well, and the code we have used (with some minor additions) is one of the most respected and robust versions freely available. For some special problems other codes might work better, but this is probably close to an universal algorithm. It uses reverse communication, which makes it rather easy to implement into programs. However, there is still room to possibly improve the update method for DFT problems because as we have mentioned, the gradient comes almost for free:
\begin{itemize}
	\item Conventional trust region methods discard a bad step; it might be better to incorporate this information into the BFGS update then recalculate a revised step. 
	\item	Most BFGS codes attempt to keep the local memory requirements and CPU time low. However, for a DFT problem these are generally negligible compared to what the main iterations require. Hence one might improve the codes by doing more detailed and accurate analysis of plausible steps; for instance, going beyond a simple double dogleg method.
	\item As mentioned above the weakness of the BFGS method when the curvature condition fails might be something where further research would be useful. For instance, one could easily keep a step history and switch to a limited-memory method or an update based upon less than the full number of steps. One idea would be to search over all the history of previous steps to find an ``optimum'' Hessian estimate that will do better than the three conventional methods described above.
\end{itemize}

\section{\label{sec:implementation}Implementation}
The general approach for constructing the initial Hessian approximation is now outlined.  First, the symmetry independent atom set is expanded by the appropriate operators to construct the full set of atoms in the structure. A search for the free variable parameters to be optimized is subsequently performed, identifying special sites whose positions are not allowed to vary for the particular symmetry of the structure. A nearest neighbor search algorithm is then carried out over the expanded atom net, to a user specified cutoff distance, in order to determine the length over which the interatomic potential acts. The cutoff terms and bonding strength used are discussed later. The elements in a trial Hessian   are then generated by numerical differentiation of a simple pairwise energy with a step size of $10^{ - 6} $
\AA, which tests show to be adequate. The final step is to construct an initial estimate to be used in the form
\[
{\mathbf{B}}_{{\text{initial}}}  = \gamma {\mathbf{B}}_{{\text{trial}}}  + \eta {\mathbf{I}}
\]
where $\gamma $ and $\eta $ are constants, and $\mathbf{I} $ is the identity matrix.

We have experimented with two models for the pairwise energy, a spring model and a simple harmonic approximation. The harmonic model consistently outperformed the spring model, for reasons which we believe are associated with the conditioning which we will discuss later. The harmonic model can be written as
\[
\Delta E \approx \tfrac{1}{2}\Gamma _{i,j} \Delta r_{i,j}^2
\]
where $\Delta {{r}}_{i,j} $ is the change in the distance between the two atoms $i$  and $j$, and $\Gamma _{i,j} $ is an appropriate spring constant linking them. Here an exponential term is used to model the pairwise bond strength, $\Gamma _{i,j}  = \exp \left[ { - \nu \left( {r_{i,j}  - d_{\text{m}} } \right)/d_{\text{m}} } \right]$, where $\nu$ is a user-specified exponential decay term (discussed later) and $d_{\text{m}} $ is the shortest nearest neighbor distance. For practical purposes the absolute value of the spring constants are not important, but only the relative ratios of them.

After building the full Hessian for the structure, it is symmetry reduced to contain only the symmetry independent atoms and transformed to conventional crystallographic fractional units. Finally, a Cholesky factorization using a LINPACK routine (\textsc{dchdc}) is performed and the Hessian approximation is introduced in the first step of a slightly modified version of the \textsc{dmng.f} minimization routine from NETLIB. This minimizer was incorporated into the geometry optimization routine found in WIEN2k by one of us (LDM) some time ago, and is now widely used.

All that is required by the user is a file from which the crystal structure is read, and a parameter file which contains constants used in the model during the Hessian construction. The parameter file was found to be quite useful, as it allows for the user to tailor the Hessian for different types of systems (e.g.~soft or hard). The parameters that were found most useful and which have been included in the model can be found in Table \ref{tab:parameters}. Values that have been shown to be quite reasonable for most calculations are also listed.

\begingroup
\squeezetable
\begin{table}
\caption{\label{tab:parameters}Free model parameters in the Hessian algorithm that may be used to customized the estimate.}
\begin{ruledtabular}
\begin{tabular}{ccl}
Parameter &	Values &	Description\\
\hline
$\chi $ &0.05				&Cutoff term limiting the number of atom pairs\\
$\nu$		&1.50-2.50	&Strength of exponential decay bonding term \\
$R_{\text{M}}$	&8-12	&Maximum nearest neighbor distance \\
$\gamma $ &0.20-0.40	&Multiplicative rescaling term \\
$\eta $ &1.00-4.00	&Additive diagonal rescaling term \\
\end{tabular}
\end{ruledtabular}
\end{table}
\endgroup

A description of each parameter follows: $R_{\text{M}}$ defines the maximum distance (in atomic units) to which the nearest neighbor search algorithm includes atoms for building the energy terms; $\nu$ is used in the exponential decay function ($\Gamma$), which describes the strength of the pairwise interaction between atoms. An additional cutoff term ($\chi$) is used to restrict which atom pairs are included in the gradient calculation. Once the exponential weighting factor becomes smaller than this $\chi$-value, those bonds (or atoms pairs) are no longer considered in the force calculation. The affect of varying these parameters is discussed later. Finally, the two scaling terms  $\gamma$ and  $\eta$.

\section{\label{sec:results}Results}
The initializer described above has been used for several months for a range of problems, and appears to behave well for cases where the initial point is both close to or far from the minimum. To investigate the effectiveness of the initializer, we used the all-electron DFT code WIEN2k and implemented it into the structure optimization routine \cite{REF:Bla2001}. In WIEN2k space within the unit cell is partitioned into spheres that define the atomic regions and an interstitial region linking them. The basis functions are then composed of atom-like wavefunctions inside the spheres centered about the ion cores with a radius ($R_{\rm MT}$) determined by the muffin-tin approximation, and outside these atomic regions by a planewave expansion. The partial solutions to the Kohn-Sham equations are then matched at the interface. For more details on WIEN2k see Ref.\ \cite{REF:Bla2001}, additionally a review of the LAPW method has been given by Singh \cite{REF:singhreview}.  To evaluate the performance in detail, we have relaxed in a more systematic fashion a series of different structures with varying degrees of freedom (relaxation in one or more Cartesian coordinate directions). A summary of these structures is given in Table \ref{tab:structures}. For these examples the PBE-GGA exchange-correlation functional \cite{REF:Per96b} was used with a planewave cutoff of RK$_{\rm max}$=7.00. Values for the muffin-tin radii used in the calculations can be found in Table \ref{tab:radii} for each system. Multiple convergence criteria were also required and they are as follows: (1) the force vector on each atom was less than 1 mRy/a.u.; (2) the energy tolerance was 0.1 mRy; and (3) the charge convergence within the muffin-tins was $5.0\times10^{-5}e$. All of the optimized energies $(E_0)$ for the structures are available in Table \ref{tab:results}. 

\begingroup
\squeezetable
\begin{table}
\caption{\label{tab:structures}Material systems investigated with relevant crystallographic information.}
\begin{ruledtabular}
\begin{tabular}{lllcc}
& & Space Group  & &\\
System &	Lattice &	(symmetry) & Atoms & D.O.F.\\
\hline
SiO$_2$   					 &primitive tetragonal& 136 (\emph{P4$_2$/mnm})& 6 & 1 \\
LaCuOS 							 &primitive tetragonal& 129 (\emph{P4/nmm})& 8 & 2\\
MgVO$_3$						 &centered orthorhombic& $\,$ 65 (\emph{Cmmm})& 10 & 3 \\
SiO$_2$							 &rhombohedral& 154 (\emph{P3$_2$21})& 9 & 4\\
Bi$_4$Ti$_3$O$_{12}$ &body centered tetragonal& 139 (\emph{I4/mmm})\footnote{Symmetry about odd digits (B2cb), where the digits are the number of TiO$_6$ octahedra in the perovskite-like fragments of the structure.} & 38& 28\\
\end{tabular}
\end{ruledtabular}
\end{table}
\endgroup

\begin{table*}
\caption{\label{tab:results}Summary of optimization results. The number of cycles are given for the default minimization method ($n_{\text{DF}}$), the Hessian initializer ($n_{\text{INIT}}$) and a converged BFGS Hessian ($n_{\text{CV}}$). Similarly, the length of the first geometry step sizes from the minimization algorithm are given for each approach. A * indicates a Cauchy step, and a \# dogleg step, while those without any denotation are of the standard Newton type. The absolute values for the starting geometry energy ($E_{s}$) and the total converged energy ($E_{0}$) are provided.}
\begin{ruledtabular}
\begin{tabular}{lcccccccc}
 & \multicolumn{3}{c}{Iterations} & \multicolumn{3}{c}{Step Size} &  \\
  \cline{2-7}
System &$n_{\text{DF}}$ &$n_{\text{INIT}}$&$n_{\text{CV}}$&$\Delta_{\text{DF}}$&$\Delta_{\text{INIT}}$&$\Delta_{\text{CV}}$& E$_s$ (Ry) & E$_0$ (Ry)  \\
\hline
SiO$_2$ &3&2&1& 0.098&0.035&0.170&62.719&62.757\\
LaCuOS   &5&5&2&0.105&0.233&0.706&9.415&9.480\\
MgVO$_3$ &10&5&5&0.153& 0.555$^*$ &0.894$^*$&2.147&2.493\\
SiO$_2$  &10&6&6&0.152&0.268&0.649$^{\#}$&27.047&27.100\\
Bi$_4$Ti$_3$O$_{12}$ &31 &21&14&0.024&0.055&0.184&58.397&58.542\\
\end{tabular}
\end{ruledtabular}
\end{table*}

\begingroup
\squeezetable
\begin{table}
\caption{\label{tab:radii}Muffin-tin radii ($R_{\rm MT}$) used for the total energy calculations in the all-electron plane wave code WIEN2k.}
\begin{ruledtabular}
\begin{tabular}{ll}
System &	$R_{\rm MT}$ (bohr) \\
\hline
SiO$_2$   					 &Si=1.70 and O=1.30\\
LaCuOS 							 &La=2.40, Cu=2.20, O=1.70 and S=2.00\\
MgVO$_3$						 &Mg=1.60, V=1.60 and O=1.40\\
SiO$_2$							 &Si=1.70 and O=1.50\\
Bi$_4$Ti$_3$O$_{12}$ &Bi=2.28, Ti=1.74 and O=1.540\\
\end{tabular}
\end{ruledtabular}
\end{table}
\endgroup

For each structure, both the number of geometry steps required by the minimization routine and the number of self-consistent field (SCF) calculations to satisfy the convergence criterion were reduced with our initialized Hessian. Table \ref{tab:results} lists the number of SCF cycles required to achieve an equilibrium structure using both a simple diagonal initializer and our Hessian formulation. Throughout the analysis we examine only the deviation in number of SCF cycles for each estimate, and note that these values are similar to the number of geometry minimization steps required. For each case examined, no more than three steps were rejected by the Trust-Region algorithm. (It is worth mentioning that the diagonal initializer was previously optimized to be close to the best, general form available, a factor of at least 30\% or more better than a simple unitary scaling.) In general the number of SCF cycles to converge was reduced by 30\% compared to the reference diagonal Hessian.

\subsection{\label{sec:level2:efficiency}Hessian estimate efficiency}
In order to evaluate the accuracy of the initial Hessian and its effects on convergence, structure relaxations were also performed using a converged BFGS Hessian from a previous calculation. We cannot prove that this is the best possible Hessian, but from previous experience it appears to be very close to optimum. Table \ref{tab:results} shows the results for the number of SCF cycles required to reach convergence with each of the different Hessians. It is important to recognize that the number of SCF cycles required with our initial estimates approached that of the converged Hessian. As the system grows in complexity (size and number) there remains room for improvement in the reduction of SCF cycles that can be obtained. The deviation between the number of SCF cycles for the converged Hessian and the initialized Hessian (in for example Bi$_4$Ti$_3$O$_{12}$) can be overcome by using a better potential to describe the pairwise bonding. In these larger structures, the Hessian is sensitive to small deviations in the off-diagonal elements, for this reason a longer-range interaction potential may show even better convergence. Nonetheless, the experiments suggest we are approaching the limit at which an equilibrium structure can be found using current optimization methods. As expected, the better the initial Hessian estimate of the true curvature of the PES, the faster the optimization. Consequently, we can conclude that our pairwise potential acting over several nearest neighbors adequately provides an estimate of the curvature of the PES. We also offer a more rigorous comparison in the next section by examining the eigenvalues of each Hessian matrix. 

We also studied the effect of the Hessian estimate on the initial geometry step size used in the BFGS update. The length of the first geometry step for each structure is given in Table \ref{tab:results}. It is clear that there is a decrease in the number of SCF cycles required with increasing maximum geometry step size. The increased reduction in geometry optimization steps (data not shown) is attributed to a more accurate Hessian (e.g. closely approximating the eigenvalues). For a smaller geometry step length the time to convergence increases, and the user can be fairly well guaranteed that the minimization will proceed stably. A more aggressive approach is to increase the geometry step size permitted in the minimization algorithm, which can be done with confidence if the initial Hessian resembles the curvature of the PES. The geometry step sizes given in Table \ref{tab:results} also suggest that our estimate is better than the standard initialization, since the initial step is much larger. It might appear then, that by taking a larger geometry step size, it is possible to reduce the number of iterations; however, this may result in the BFGS update moving in directions of higher energy at first, before final convergence is achieved. Increasing the step size too aggressively may therefore result in more steps than desired.

\subsection{\label{sec:level2:conditioning}Hessian conditioning}

It is known that the rate of minimization for steepest decent and conjugate methods is related to the condition number   of the Hessian matrix. The condition number is defined as the ratio of ${{\omega _{\max } } \mathord{\left/ {\vphantom {{\omega _{\max } } {\omega _{\min } }}} \right. \kern-\nulldelimiterspace}{\omega _{\min } }}$, where $\omega _{\max }$ and $\omega _{\min }$  are the largest and smallest eigenvalues, respectively. Typically, the minimizations steps required scales with the condition number \cite{nocedal:wright:1999}. Essentially, the condition number of a matrix measures how sensitive its inverse is in changes to the original matrix: for a large condition number the inverse of the matrix is very sensitive or unstable; a matrix with a low condition number (bounded by unity) is said to be well-conditioned, while a matrix with a high condition number is said to be ill-conditioned. From our experience, it turned out to be important to consider the conditioning of the initial Hessian.

The results in Table \ref{tab:results} are from appropriately conditioned ${\mathbf{B}}_{{\text{initial}}} $   matrices, whose properties (condition numbers and eigenvalues) are given in Table \ref{tab:eigs}. The calculation of condition numbers, eigenvalues and their corresponding eigenvectors were performed with standard LAPACK routines for real symmetric matrices \cite{LAPACK:UG}.

\begingroup
\squeezetable
\begin{table}
\caption{\label{tab:eigs}Optimal scaling values for each structure and the largest and smallest eigenvalues along with the condition number for each model Hessian matrix. The remaining parameters were fixed at $R_{\text{M}}= 10.0$, $\nu  = 2.00$, and $\chi  = 0.05$ for structure.}
\begin{ruledtabular}
\begin{tabular}{lcccc}
System &	$\gamma$ & $\eta$ & $\omega _{\min },\omega _{\max }$ & $\kappa$ \\
\hline
SiO$_2$   					 &2.75&1.05&1.0500, 12.050&11.476\\
LaCuOS 							 &0.15&1.05&1.3179, 1.3821&1.0487\\
MgVO$_3$						 &0.50&1.05&1.7298, 2.5868&1.4954\\
SiO$_2$							 &0.25&1.05&1.2998, 1.8722&1.4404\\
Bi$_4$Ti$_3$O$_{12}$ &0.12&1.05&1.0500, 1.4129&1.3457\\
\end{tabular}
\end{ruledtabular}
\end{table}
\endgroup

\begin{figure*}
	\centering
		\includegraphics[width=0.98\textwidth]{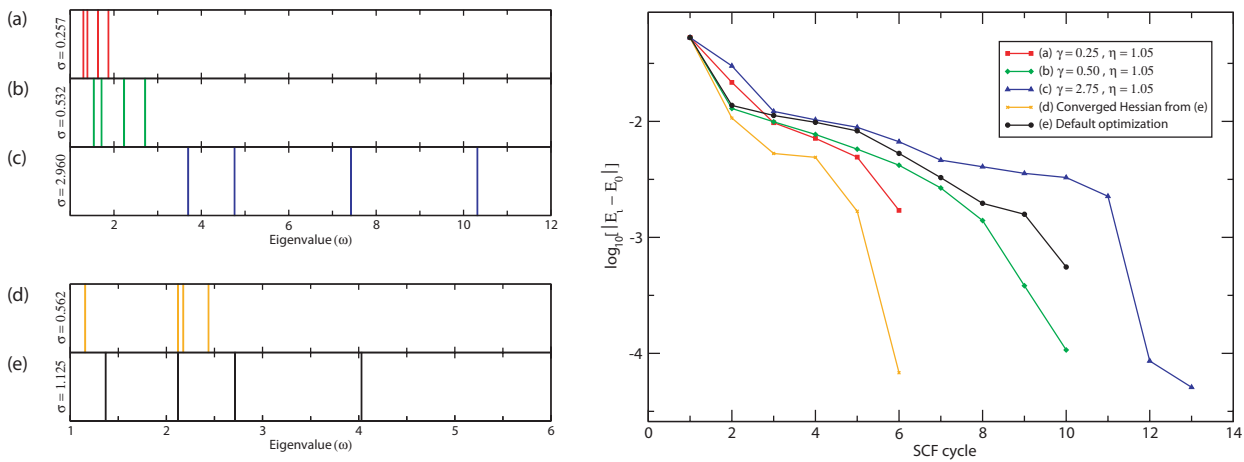}
		\caption{\label{fig:sio2}(color online)~Left panels: Eigenvalue distribution of the Hessian matrix for the rhombohedral SiO$_2$ structure with different Hessian scaling values (a) $\gamma$ = 0.25, $\eta$ = 1.05; (b) $\gamma$ = 0.50, $\eta$ = 1.05; (c) $\gamma$ = 2.75, $\eta$ = 1.05; (d) Converged BFGS Hessian; and (e) the default optimization. The standard deviation ($\sigma$) for the eigenvalues is given for each case. Right panel: Convergence of the total energy as a function of the SCF cycle number for each calculation. $E_{0}$ is the converged value of the total energy for each run.}
\end{figure*}

\begin{figure*}
	\centering
		\includegraphics[width=0.98\textwidth]{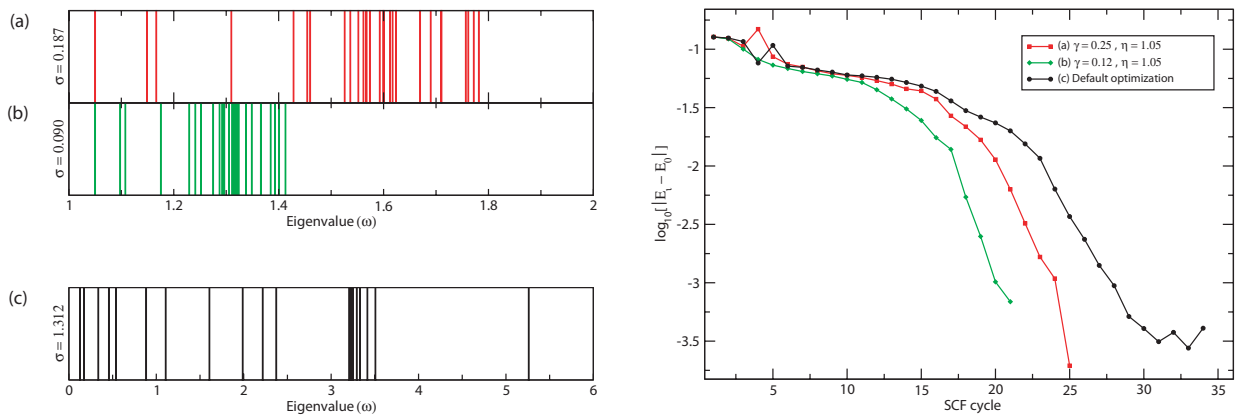}
		\caption{\label{fig:bto}(color online)~Left panels: Eigenvalue distribution of the Hessian matrix for the Bi$_4$Ti$_3$O$_{12}$ structure with different Hessian scaling values (a) $\gamma$ = 0.25, $\eta$ = 1.05; (b) $\gamma$ = 0.50, $\eta$ = 1.05; and (c) the default optimization. The standard deviation ($\sigma$) for the eigenvalues is given for each case. Right panel: Convergence of the total energy as a function of the SCF cycle number for each calculation. $E_{0}$ is the converged value of the total energy for each run.}
\end{figure*}

To explore the scaling effect, we present more detailed results for the rhombohedral SiO$_2$ and Bi$_4$Ti$_3$O$_{12}$ structures. Similar analyses were done on the other structures examined, and the results presented are representative of the general trends. Figs.~\ref{fig:sio2} and \ref{fig:bto} show the convergence in energy for various scaling parameters. While the condition number of the Hessian is a good estimate at how successful the optimization will be, we have found that a better metric is to examine the eigenvalues of the Hessian matrix. The eigenvalue distributions are shown in the left panels of Figs.~\ref{fig:sio2} and \ref{fig:bto} for SiO$_2$ and Bi$_4$Ti$_3$O$_{12}$. We find that an average condition number is best (see for example \cite{nash}), and a tight cluster of the eigenvalues (small standard deviation) is desired. From these figures, it is clear that with a wider distribution of the Hessian eigenvalues, the structure relaxation performance declines. Additionally, the minimization occurs more stably when the eigenvalue distribution is narrow. 

Our study suggests that we want to minimize the ratio between the largest and smallest eigenvalues (as expected) to optimize the condition number. However, we do not want the matrix overly conditioned. In fact, the most robust geometry relaxation occurs when the eigenvalues deviate slightly from those of the true curvature. Rather than achieving a fully converged Hessian (i.e. the eigenvalues correctly replicate the true PES) at a point in space far from the optimal geometry, it is better to have the eigenvalues gradually converge toward the true values as the system moves from the initial geometry to the equilibrium configuration as suggested by Olsen \emph{et al.} \cite{olsen:9776}. In fact, the BFGS method can achieve convergence without actually replicating the true Hessian of PES.

\section{\label{sec:discussion}Discussion}
Throughout this investigation we have attempted to reach the optimal rate of geometry relaxation for a given structure.  We have reached several conclusions on the optimal values of the scaling parameters, and an approach to finding them. We note that in optimization problems, scaling tends to be the responsibility of the investigator owing to the variety of applications; however, at times it is not always clear what is the best or most robust means of doing so. 

This study has allowed us to build a set of parameters that can aid in the relaxation of large condensed structures. These parameters listed in Table \ref{tab:parameters} are in order of increasing importance and their effects on convergence are now discussed. The most straightforward parameters to set are $\chi$, $\nu$ and $R_{\text{M}}$. From our experience, variation in these parameters only affected optimization performance by on average a few geometry steps (and no change in the number of SCF cycles) over the listed range. The most significant effect of these parameters was found to be on the initial step size in the BFGS update. Most notably, increasing $\nu$ resulted in larger steps sizes by a few percent.

In order to further optimize the efficiency of the minimization algorithm, scaling of the Hessian through the $\gamma$  and $\eta$  parameters was investigated. The best ${\mathbf{B}}_{{\text{initial}}} $ matrix seems to be a balance between appropriately scaling the diagonal elements with respect to the off-diagonal elements and eigenvalue clustering. As we have shown, intelligent choices for the scale parameters can enhance performance. We note that in our model as the ratio of ${\gamma  \mathord{\left/ {\vphantom {\gamma  \eta }} \right. \kern-\nulldelimiterspace} \eta }$ increases the condition number of the Hessian increases exponentially.

We have also found that our conditioning method has led to an increase in the number of Newton-type steps in the minimization algorithm. This fact suggests that the algorithm may be behaving more like metric based minimization techniques, i.e.\ Newton methods where adjacent steps are forced to be conjugant to each other $\left( {{\mathbf{s}}_{k + 1} {\mathbf{B}}_k {\mathbf{s}}_k  = 0} \right)$. In addition, the convergence dependency on the eigenvalue structure has been know for sometime for conjugate gradient methods \cite{axelsson,notay}, but this behavior for BFGS methods is less well-documented. In fact, the eigenvalue distribution has been reported to have minimal influence on convergence rates through tests performed on large molecules \cite{Baysal}. However, as the object function more accurately describes the PES, i.e.\ as harmonic about the minimum, the BFGS update behaves like a variable metric method in the sense that steps of optimal length are chosen. Furthermore, the eigenvalue cluster phenomenon is consistent  with  quasi-Newton type updating as Morales and others have shown \cite{morales}. The eigenvalue spectrum of a matrix can then be preconditioned to form narrow clusters in order to accelerate convergence in optimization problems. The preconditioning effects are a general property for solutions to linear systems and for that reason are  scalable. Our results may then be applied to preconditioning a large set of structures requiring optimization.

While generalized minimization algorithms are necessary foundations for structure calculations, tailoring of the geometry relaxation routines provides a robust means for enhancing performance. Of course, the optimal approach will be different for different structures. For practical purposes, it is important that these geometry relaxations be run with very little parameter adjustments by the user. Therefore from the previous considerations, only variations in the parameters which affect the conditioning of the matrix and the  clustering of the eigenvalues should be considered, i.e.~reduce the standard deviation in the eigenvalues of ${\mathbf{B}}_{{\text{initial}}} $. The effect of this clustering seems to be a consequence of the optimization method (the BFGS update) and is still being investigated. 

\section{conclusion}
We have developed a customizable model (any interaction potential can be substituted) which adequately approximates the curvature of the potential energy surface of a crystal structure. The model has been parameterized to allow for modification for different system types. We have shown that our method results an approximate 30\% decrease in the number of SCF cycles required to achieve an equilibrium structure relative to the standard routines. In fact, our estimate is shown to closely replicate the behavior of a converged BFGS Hessian. The effects of preconditioning have also been investigated, and a general approach for enhancing the rate of convergence through scaling factors has been suggested.

\begin{acknowledgments}
We wish to acknowledge D.~Russell Luke for helpful discussions on optimization methods and scaling techniques. Peter Blaha also provided the Bi$_4$Ti$_3$O$_{12}$ structure and assisted in the beta-testing of the \textsc{pairhess} program. This work was funded by the NSF under  GRANT \# DMR-0455371 
\end{acknowledgments}

\bibliography{hessBIB}

\end{document}